\documentclass[sigconf]{acmart}

\AtBeginDocument{%
  \providecommand\BibTeX{{%
    \normalfont B\kern-0.5em{\scshape i\kern-0.25em b}\kern-0.8em\TeX}}}

\copyrightyear{2021} 
\acmYear{2021} 
\setcopyright{acmcopyright}
\acmConference[MM '21]{Proceedings of the 29th ACM International Conference on Multimedia}{October 20--24, 2021}{Virtual Event, China}
\acmBooktitle{Proceedings of the 29th ACM International Conference on Multimedia (MM '21), October 20--24, 2021, Virtual Event, China}
\acmPrice{15.00}
\acmDOI{10.1145/3474085.3481548}
\acmISBN{978-1-4503-8651-7/21/10}

\usepackage{graphicx}
\usepackage{subfigure}
\usepackage{multirow}
\usepackage{bm, amsmath}
\usepackage{makecell}
\usepackage{color}
\usepackage{rotating}
\usepackage{balance}
\usepackage{enumitem}

\begin{document}
\fancyhead{}

%%
%% The "title" command has an optional parameter,
%% allowing the author to define a "short title" to be used in page headers.
\title[Improving Fake News Detection by Using an Entity-enhanced Framework to Fuse Diverse Multimodal Clues]{Improving Fake News Detection by Using an Entity-enhanced Framework to Fuse Diverse Multimodal Clues}

%%
%% The "author" command and its associated commands are used to define
%% the authors and their affiliations.
%% Of note is the shared affiliation of the first two authors, and the
%% "authornote" and "authornotemark" commands
%% used to denote shared contribution to the research.

\author{
Peng Qi\textsuperscript{1,2,3}, %$^{1,2}$
Juan Cao\textsuperscript{1,2}, 
Xirong Li\textsuperscript{4}, 
Huan Liu\textsuperscript{5}, 
Qiang Sheng\textsuperscript{1,2},
Xiaoyue Mi\textsuperscript{1,2}, \\
Qin He\textsuperscript{6}, 
Yongbiao Lv\textsuperscript{6}, 
Chenyang Guo\textsuperscript{6}, 
Yingchao Yu\textsuperscript{6}
}
% 1 ict, 2 ucas, 3 hebi, 4 renda, 5 zhengzhou 6, ruijian
\affiliation{%
  \institution{
$^{1}$Key Lab of Intelligent Information Processing, Institute of Computing Technology, CAS, Beijing, China\\
$^{2}$University of Chinese Academy of Sciences
$^{3}$Institute of Artificial Intelligence, Hebi, China\\
$^{4}$Key Lab of DEKE, Renmin University of China, Beijing, China
$^{5}$Zhengzhou University, Zhengzhou, China\\
$^{6}$Hangzhou ZhongkeRuijian Technology Co., Ltd., Hangzhou, China\\
}
\country{}
}

\email{{qipeng,caojuan,shengqiang18z,mixiaoyue19s}@ict.ac.cn,xirong@ruc.edu.cn,liuhuan_2012@hotmail.com,}
\email{{heqin,lvyongbiao,guochenyang, yuyingchao}@ruijianai.com}

%%
%% By default, the full list of authors will be used in the page
%% headers. Often, this list is too long, and will overlap
%% other information printed in the page headers. This command allows
%% the author to define a more concise list
%% of authors' names for this purpose.
%\renewcommand{\shortauthors}{Trovato and Tobin, et al.}

%%
%% The abstract is a short summary of the work to be presented in the
%% article.
\begin{abstract}
Recently, fake news with text and images have achieved more effective diffusion than text-only fake news, raising a severe issue of multimodal fake news detection. 
Current studies on this issue have made significant contributions to developing multimodal models, but they are defective in modeling the multimodal content sufficiently.
Most of them only preliminarily model the basic semantics of the images as a supplement to the text, which limits their performance on detection. 
In this paper, we find three valuable text-image correlations in multimodal fake news: entity inconsistency, mutual enhancement, and text complementation. 
To effectively capture these multimodal clues, we innovatively extract visual entities (such as celebrities and landmarks) to understand the news-related high-level semantics of images, and then model the multimodal entity inconsistency and mutual enhancement with the help of visual entities. Moreover, we extract the embedded text in images as the complementation of the original text. 
All things considered, we propose a novel entity-enhanced multimodal fusion framework, which simultaneously models three cross-modal correlations to detect diverse multimodal fake news. 
Extensive experiments demonstrate the superiority of our model compared to the state of the art. 
\end{abstract}

%%
%% The code below is generated by the tool at http://dl.acm.org/ccs.cfm.
%% Please copy and paste the code instead of the example below.
%%


%%
%% Keywords. The author(s) should pick words that accurately describe
%% the work being presented. Separate the keywords with commas.

%%
%% This command processes the author and affiliation and title
%% information and builds the first part of the formatted document.
\maketitle

%{%\par
%  \medskip\small\noindent{\bfseries ACM Reference Format:}\par\nobreak
%  \noindent\bgroup\def\\{\unskip{}, \ignorespaces}{Mingxi Zhang, Xifeng Yan, Wei Wang}\egroup. 2021. Comprehensively Computing Link-based Similarities by Building A Random Surfer Graph. In \textit{Proceedings of the 30th ACM Int'l Conf. on Information and Knowledge Management (CIKM '21), November 1--5, 2021, Virtual Event, Australia}\textit{.} ACM, New York, NY, USA, \ref{TotPages}~pages. https://doi.org/10.1145/XXXXXX.XXXXXX
%  }

\section{Introduction}

The rising prevalence of fake news and its alarming real-world impacts have motivated both academia and industry to develop automatic methods to detect fake news (i.e., designing a classifier to judge a piece of given news as real or fake) \cite{surveykdd2017, surveycsur2018, surveykumar2018false, surveykdd2019, surveycsur2020}. Traditional approaches \cite{textCastillo, majingijcai, texticcl2018, textemnlp2011} typically focus on the textual content, which is the main description form of news events. 
With the recent evolution of fake news from text-only posts to multimedia posts with images or videos \cite{chapter}, approaches based on multimodal content demonstrate promising detection performance \cite{attRNN, eann, mvae, mkemn, spotfake}. This paper targets multimodal fake news detection, which is utilizing information of multiple modalities (here, text and images) to detect fake news. 

\begin{figure}
	\centering
	\begin{minipage}[b]{0.5\textwidth}
		\subfigure[Entity Inconsistenty]{
	\includegraphics[width=0.49\textwidth]{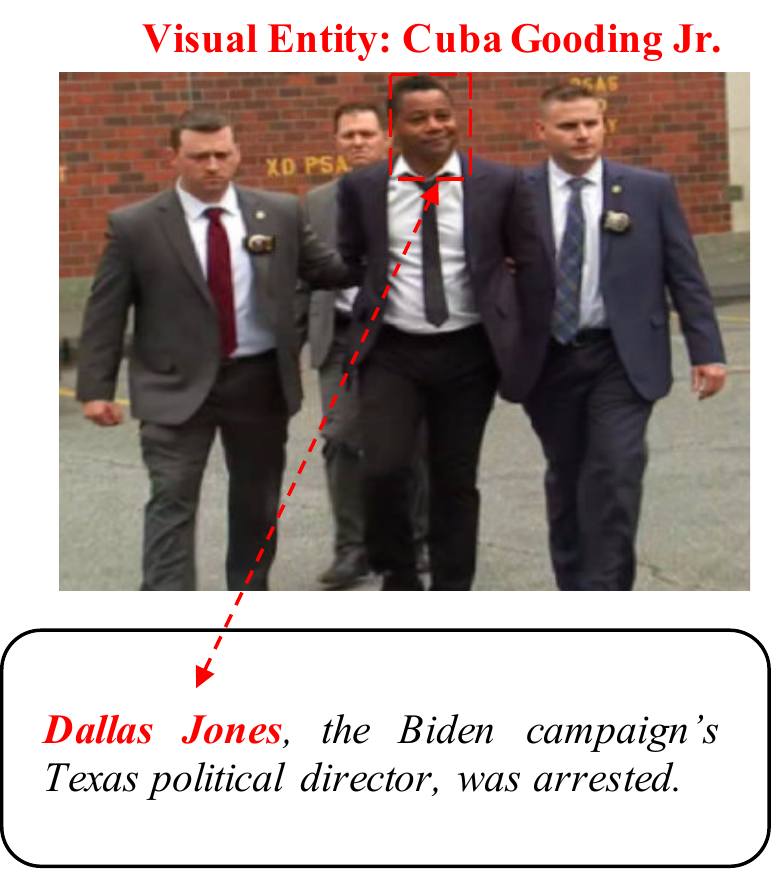}
			\label{fig:outdated}
	}
	\subfigure[Mutual Enhancement]{
			\includegraphics[width=0.49\textwidth]{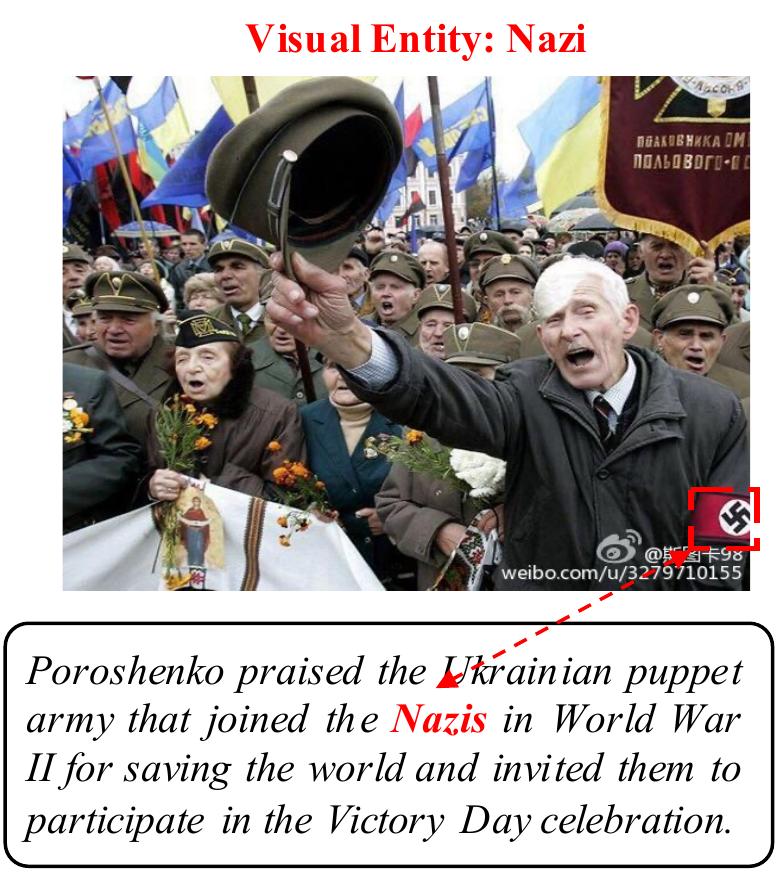}
			\label{fig:decorative}
	}
	\end{minipage}
	\\
	\begin{minipage}[b]{0.5\textwidth}
		\centering
		\subfigure[Text Complementation]{
			\includegraphics[width=0.64\textwidth]{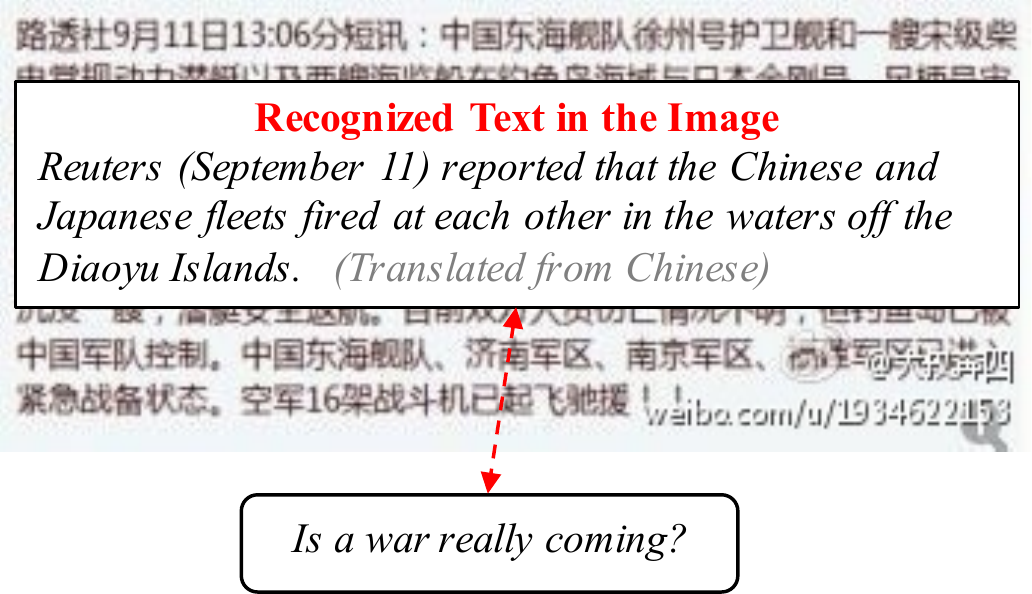} 
			\label{fig:text-embedded}
	}
	\end{minipage}
	\caption{Three valuable text-image correlations in multimodal fake news, which provide diverse clues for detection.}
	\label{fig:introcase}
\end{figure}

\begin{figure*}
\includegraphics[width=.95\textwidth]{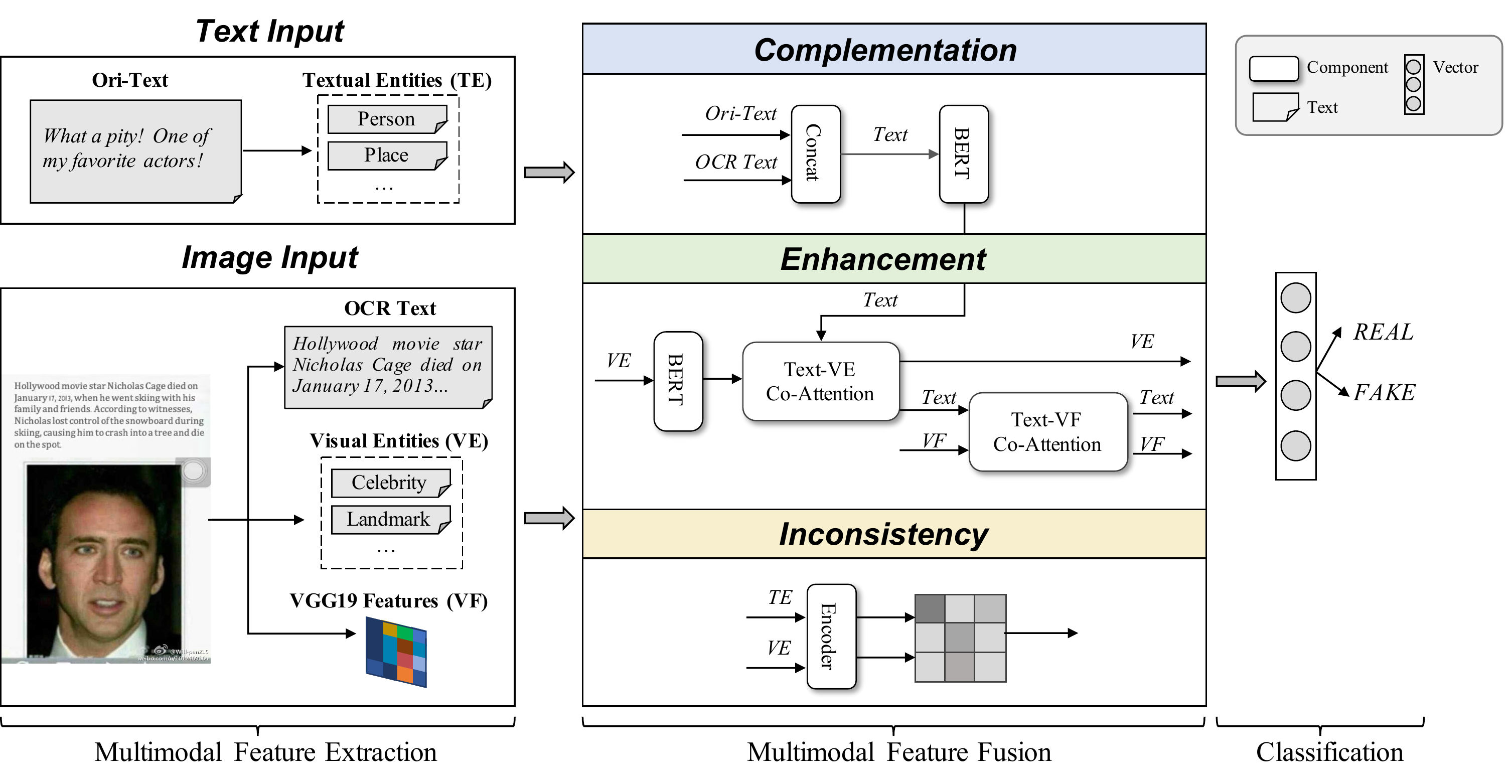}
	\caption{Architecture of the proposed framework EM-FEND. 
	In the stage of \textit{Multimodal Feature Extraction}, we explicitly extract the textual and visual entities to model the key news elements, and extract the OCR text and visual CNN features of the input image. 
	In the stage of \textit{Multimodal Feature Fusion}, we model three text-image correlations, that is text complementation, mutual enhancement, and entity inconsistency.  
	Finally, these multimodal features are fused by concatenation for \textit{Classification}. }
	\label{fig:model}
\end{figure*}

Despite recent advancements in developing multimodal models to detect fake news, 
existing works model the multimodal content insufficiently. 
Most of them only preliminarily model the basic semantics of the images as the complement of the text, ignoring the characteristics of multimodal fake news. 
Specifically, some prior arts \cite{eann, spotfake, metaFEND} obtain the multimodal representations by simply concatenating the textual features with visual features extracted from VGG19 \cite{vgg19} that is pre-trained on ImageNet \cite{imagenet}.
 
To make up for this omission, we explore three valuable text-image correlations in multimodal fake news, which provide diverse multimodal clues. 
a) \textbf{Text and images have inconsistent entities}, which is a potential indicator for multimodal fake news. Wrongly reposting outdated images is a typical way to make up multimodal fake news \cite{mediaeval15, mediaeval16, mvnn}. However, it is difficult to find both semantically pertinent and non-manipulated images to support these non-factual stories in fake news, causing the inconsistency between text and images. For example, as shown in Figure ~\ref{fig:outdated}, the text describes a piece of news about "Dallas Jones" while the attached image is the arrest scene of another person.   
%This kind of fake news is image-centered, of which the images provide vital clues for detection. 
b) \textbf{Text and images enhance each other by spotting the important features.} News text and images are related in high-level semantics, and the aligned parts usually reflect the key elements of news. 
In this kind of multimodal fake news, the text provides main clues for detection, while images help select the key clues in the text. 
As Figure ~\ref{fig:decorative} shows, the  Nazi flag in the image corresponds to the important entity "Nazi" in the text, which is the key controversial point of this news post. 
%In this fake news, important news entities mentioned in the text could be illustrated and emphasized by images and vice versa.   
c) \textbf{The embedded text in images provides complementary information for the original text. } According to our preliminary statistics on the Weibo dataset \cite{attRNN}, more than 20\% of multimodal fake news spreads in the form of image. This refers to news that the embedded text in the image tells the complete fake news story while the original text often is comment (see Figure ~\ref{fig:text-embedded}). In this kind of fake news, the clues lie in the combination of the original text and the embedded text in the image.   

In addition to the diversity of multimodal clues, 
another challenge of fusing multimodal information for detection lies in the heterogeneity of multimodal data. Current works focus on the general objects of news images by pre-trained VGG19 or Faster R-CNN, while the news text is in a more abstract semantic level based on named entities\footnote{A narrow definition of name entities are objects that can be denoted with a proper name
such as persons, organizations, and places \cite{namedentity}.}.
%"phrases that contain the names of persons, organizations and location" \cite{namedentity}.}.s
 Due to this semantic gap, current works are hard to reason effectively between text and images for exploiting multimodal clues. For example, as shown in Figure ~\ref{fig:outdated}, we can't reveal the multimodal inconsistency as clues to detect this news as fake if we only recognize the celebrity in the image as "person" instead of "Cuba Gooding Jr.".

To address this challenge, we innovatively import the visual entities to model the high-level semantics of news images. The visual entities consist of words describing named entities recognized from the images (such as celebrity and landmark) and some news-related visual concepts. 
They are important for mining the multimodal clues because they 1) contain rich visual semantics and thus help understand the multimodal news, and 2) bridge the high-level semantic correlations of news text and images.

All things considered, we propose a novel framework for multimodal fake news detection, named as \textbf{EM-FEND} (\textbf{E}ntity-enhanced \textbf{M}ultimodal \textbf{F}ak\textbf{E} \textbf{N}ews \textbf{D}etection)
(shown in Figure ~\ref{fig:model}),
which fuses diverse multimodal clues to detect multimodal fake news. 
Specifically, 
1) in the stage of \textit{Multimodal Feature Extraction}, in addition to extract the basic visual features through fine-tuned VGG19, we explicitly extract visual entities and the embedded text in images to model the high-level visual semantics. 
Besides, we explicitly extract textual entities to capture the key elements of news events.
2) In the stage of \textit{Multimodal Feature Fusion}, we model three types of cross-modal correlations in multimodal fake news to fuse diverse multimodal clues for detection.
First, 
to model the \underline{text complementation}, we concatenate the original text and the OCR text in images as the composed text and feed it into BERT to obtain the fused textual features. 
Second, we use co-attention transformers between textual features with visual entities and visual CNN features to model the multimodal \underline{mutual enhancement} at different visual semantic levels.
Third,  
we measure the multimodal \underline{entity inconsistency} by calculating the similarity of textual and visual entities.
 And then, we fuse the above multimodal features by concatenation.      
3) In the stage of \textit{Classification}, the fused multimodal features are used to distinguish the fake and real news.  
Our main contributions are summarized as follows:
\begin{itemize}[leftmargin=17pt]
\item{We find three valuable text-image correlations in multimodal fake news, and propose a unified framework to fuse these multimodal clues simultaneously. 
}
\item{To our best knowledge, we are the first to import the visual entities into multimodal fake news detection, which helps to understand the news-related high-level semantics of images and bridge the high-level semantic correlations of news text and images.}
\item{Both offline and online evaluations demonstrate the superiority of our model compared to the state of the art. }
\end{itemize}

\section{Related Work} 
\label{relatedwork}

We will briefly review existing works on multimodal fake news detection (see Table ~\ref{tab:comparison}) and explain our novelties accordingly. 
 
The commonly used multimodal fusion framework for detection is to extract general visual features from pre-trained VGG19 \cite{vgg19} and then simply concatenate them with textual features. 
Based on this framework, Wang et al. \cite{eann} imported the event classification as an auxiliary task of fake news classification to guide the learning of event-invariant multimodal features for better generalizability.
Then, Wang et al. \cite{metaFEND} proposed a meta neural process approach to detect fake news on emergent events. 
Dhruv et al. \cite{mvae} revised this framework into a multimodal variational autoencoder to learn a shared representation of multimodal contents for classification.
Singhal et al. \cite{spotfake} first imported pre-trained language models (that is BERT, here) into this multimodal framework.
Despite the advancements made by these works, they ignore the complex cross-modal correlations in fake news, which limits the effectiveness of multimodal content in detection.

Wrongly reposting irrelevant images is a typical way to make up multimodal fake news, and thus some works focus on measuring the multimodal consistency for detection.  
Zhou et al. \cite{safe} used the image captioning model to translate the images into sentences and then computed the multimodal inconsistency by measuring the sentence similarity between the original text and the generated image captions. However, the translation performance is limited by the discrepancy between the training corpus of the image captioning model and the real-world news corpus, which further impairs the performance of cross-modal consistency measurement.
Xue et al. \cite{mcnn} transformed the textual and visual features into a common feature space by weight sharing and then computed the cosine similarity of transformed multimodal features. 
Nevertheless, it is still hard to capture the multimodal inconsistency because of the semantic gap between textual and visual features. 

On the other hand, some researchers proposed well-designed methods to model multimodal mutual enhancement. 
Jin et al. \cite{attRNN} proposed a neuron-level attention mechanism, and Zhang et al. \cite{mkemn} used the attention mechanism and multi-channel CNN to fuse multimodal information. These two works focus on the unidirectional enhancement of multimodal content, that is, highlighting the important image regions under textual guidance. Further, Song et al. \cite{ipm-song} utilized the co-attention transformer to model the bidirectional enhancement between text and images. 
Wang et al. \cite{kmgcn} extracted objects of the images and then use GCN to model the correlation between words and object labels. 
Similarly, Li et al. \cite{emaf} extracted objects and then used the Capsule network to fuse the nouns and visual features of these objects.  
Nevertheless, these methods ignore the cross-modal enhancement on high-level semantics.  
 
To sum up, there are two main drawbacks of existing works: 1) They do not consider these three cross-modal correlations simultaneously, and totally ignore the text complementation between the original text and the embedded text, and 2) model the cross-modal correlations based on the basic semantic features of the images, ignoring the news-related high-level visual semantics.    
To address these issues, we explicitly extract the visual entities and model the multimodal inconsistency and enhancement based on the multimodal entities. Moreover, we extract the embedded text in the images and model the text complementation. 
 All things considered, we design a unified framework to fuse these multimodal clues for detection.

\begin{table*}
%	[htbp]
	\caption{Comparison between EM-FEND and the state of the art for multimodal fake news detection. 
	These compared methods do not consider three cross-modal correlations at the same time.}
	\begin{center}
		\begin{tabular}{llllccc}
			\hline	
\multirow{2}{*}{\hspace{1em}\textbf{Methods}} & \multicolumn{2}{c}{\textbf{Backbone}}& &\multicolumn{3}{c}{\textbf{Cross-modal Correlations}} \\
\cmidrule{2-4}
\cmidrule{5-7}
 & \textbf{Text} & \textbf{Image} & \textbf{Fusion} & \textit{inconsistency} & \textit{enhancement}& \textit{text complementation}  \\	
 \hline
EANN\cite{eann}&Text-CNN&VGG19&concat&-&-&-\\
metaFEND\cite{metaFEND}&Text-CNN&VGG19&concat&-&-&-\\
MVAE\cite{mvae}&Bi-LSTM&VGG19&variational autoencoder&-&-&-\\
SpotFake\cite{spotfake}&BERT&VGG19&concat&-&-&-\\
\hline
SAFE\cite{safe}&Text-CNN&\makecell[l]{image2sentence\\+Text-CNN}&concat+multi-loss&text-imagecaption&-&-\\
MCNN\cite{mcnn}&\makecell[l]{BERT\\+Bi-GRU}&\makecell[l]{ResNet50\\+Attention}&attention+multi-loss&text-visfea&-&-\\
\hline
attRNN\cite{attRNN}&Bi-LSTM&VGG19&neuron-level attention&-&text->visfea&-\\
MKEMN\cite{mkemn}&Bi-GRU&VGG19&\makecell[l]{attention\\+multi-channel CNN}&-&text->visfea&-\\
CARMN\cite{ipm-song}&BERT&VGG19&\makecell[l]{co-attention transformer\\+multi-channel CNN}&-&text<->visfea&-\\
KMGCN\cite{kmgcn}&-&YOLOv3&GCN&-&text<->objects&-\\
EMAF\cite{emaf}&BERT&Faster-RCNN&Capsule&-&text<->object fea&-\\
\hline
\textbf{EM-FEND(ours)}&BERT&\makecell[l]{VGG19\\+entity detector\\+OCR model}&co-attention transformer&text-visentity&\makecell[c]{ text<->visfea\\text<->visentity}&+\\
\hline
		\end{tabular}
		\label{tab:comparison}
	\end{center}
\end{table*}

\section{Entity-enhanced Multimodal Fake News Detection}
\subsection{Model Overview}
The goal of the proposed EM-FEND framework is to predict whether the given news is real or fake by utilizing its text $T$ and the attached image $I$\footnote{Our model is applicable to news that contains multiple images, but for simplification we assume that only a single image is present in a piece of news.}.  
As shown in Figure ~\ref{fig:model}, EM-FEND includes three modules to fuse diverse multimodal clues for fake news detection: 
1) Multimodal feature extraction, which extracts the textual and visual entities, the embedded text in the image, and the visual CNN features (Section \ref{feature});
2) Multimodal feature fusion, which models three types of cross-modal correlations, including entity inconsistency, mutual enhancement, and complementation (Section \ref{fusion}); 
and 3) Classification, which uses the obtained multimodal representation to perform binary classification (Section \ref{classification}). 
We will introduce the above modules in detail. 

\subsection{Multimodal Feature Extraction}
\label{feature}
\subsubsection{Text Input}
\ 

\textbf{Textual Entities.}
As a special narrative style, news usually contains named entities such as persons and locations. These entities are of importance in understanding the news semantics and also helpful in detecting fake news. Thus, we explicitly extract the person entities $\bm{P}_T$ and location entities $\bm{L}_T$ by recognizing corresponding proper nouns in the text. 
For better understanding the news events, we employ the part-of-speech tagging to extract all nouns as a general textual context $\bm{C}_T$.   

\subsubsection{Image Input}
\ 

\textbf{Visual CNN Features.}
Following previous works, we adopt VGG19 to extract the visual features. 
Unlike these works, we fine-tune the pre-trained VGG19 on the given dataset to flexibly capture the low-level characteristics of the images from the specific data source to help detection. 
For example, the image quality is a powerful feature for distinguishing fake news and real news posts on social media, while it is less effective for detecting fake news articles on formal news sites. Then, we extract the visual features of the input image from the output of the last layer of VGG19. Considering that different regions in the image may show different patterns, we split the original image into $7 \times 7$ regions, and then obtain the corresponding visual features sequence $\bm{H}_V=[\bm{r}_1,..., \bm{r}_n], n=49$, where $\bm{r}_i$ represents the feature of the $i$-th region in the image.

\textbf{Visual Entities.}
Similar to the text, news images also contain newsworthy visual entities, which are important for semantic understanding and fake news detection.
Specifically, we extract four types of visual entities:
1) celebrities and landmarks;
2) organizations, such as Nazi, Buddhism and police, by detecting flags or clothes;
3) eye-striking visual concepts, such as violence, bloodiness, and disaster \cite{mvnn};
and 4) general objects and scenes.  
Due to the high accuracy requirements for pre-trained models and the lacking of relevant publicly available datasets, we use public APIs\footnote{https://ai.baidu.com/tech/imagecensoring, https://ai.baidu.com/tech/imagerecognition} to detect visual entities instead of re-implement these models. 
Finally, we obtain the person entities $\bm{P}_V$, location entities $\bm{L}_V$, and other news-related visual concepts with corresponding probability as a more general image context $\bm{C}_V$. 

\textbf{Embedded Text.}
In addition to the original input text, text embedded in images is also important because it usually contains important information missed by the original text. 
We extract the embedded text $O$ of the input image by applying the optical character recognition (OCR) model\footnote{https://ai.baidu.com/tech/ocr/general}.

\subsection{Multimodal Feature Fusion}
\label{fusion}

\subsubsection{Text Complementation} 
As the main body of multimodal news, the text provides rich clues for the judgment of news credibility.
For fake news in social media, in addition to the original text,
the embedded text in images is also important in understanding the news semantics and providing clues for detection.
In many situations, the key clues for detection lie in the embedded text, while the original text is just a comment about the news event.
Therefore, the original and the embedded text should be modeled jointly to obtain the whole semantics of the news event.
Most existing methods use recurrent or convolutional neural networks to model the contextual information of the textual sequence. Recently, pre-trained language models have shown strong ability in modeling text.
Thus, we feed the original text $T$ and embedded text $O$ into the pre-trained BERT \cite{bert} separated by $[SEP]$, that is
\begin{equation}
	\bm{H}_T = {\rm BERT}([CLS] T [SEP] O [SEP]).
\end{equation}
Then we obtain the textual feature $\bm{H}_T=[\bm{w}_1,..., \bm{w}_n]$, where $\bm{w}_i$ represents the feature of the $i$-th word in the composed text and $n$ is the length of the composed text. 

\begin{figure}
	\includegraphics[width=.3\textwidth]{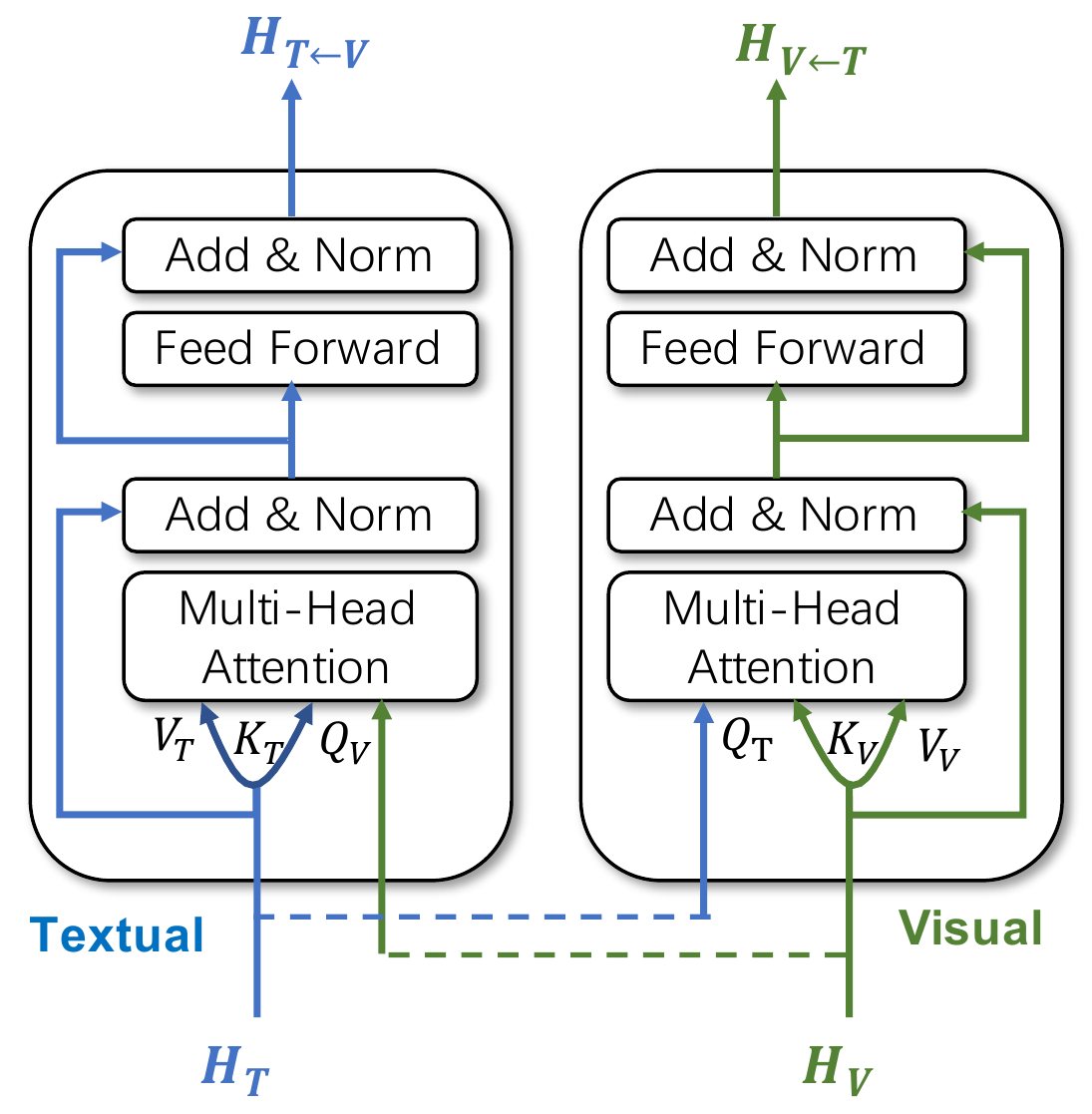}
	\caption{Multimodal co-attention transformer layer.}
	\label{fig:coatt}
\end{figure}

\subsubsection{Mutual Enhancement}
\  
In multimodal news, important news elements mentioned in the text are usually illustrated and emphasized by images and vice versa.
Thus, the text and images could spot the important features respectively by aligning with each other. Inspired by the successes of the co-attention mechanism in VQA tasks \cite{vilbert,2coattention}, we use the multimodal co-attention transformer between textual features with visual entities and visual CNN features to model multimodal alignment at different visual levels. 

\textbf{Multimodal Co-attention Transformer (MCT).}
As shown in Figure ~\ref{fig:coatt}, we use a two-stream transformer to process the textual and visual information simultaneously, and modify the standard query-conditioned key-value attention mechanism \cite{transformer} to develop a multimodal co-attentional transformer module.
The queries from each modality are passed to the other modality's multi-headed attention block, and consequentially this transformer layer produces image-enhanced textual features and text-enhanced visual features.

\textbf{MCT between Textual Features and Visual Entities.}After obtaining the visual entities $\bm{VE}=[\bm{P}_V, \bm{L}_V, \bm{C}_V]$, we employ pre-trained BERT to obtain their embeddings $\bm{H}_{VE}$. 
And thus, the textual features and visual entities' embeddings could be fused in similar BERT-constructed feature spaces, alleviating the problem of multimodal feature heterogeneity. The aligned words and visual entities usually reflect the key elements of the news, and thus we use the multimodal co-attention transformer to fuse these features. Specifically, we feed the textual features $\bm{H}_T$ and the visual entities features $\bm{H}_{VE}$ into the first co-attention transformer in Figure ~\ref{fig:model}, obtaining the textual representation enhanced by visual entities $\bm{H}_{T\gets VE}$ and text-enhanced visual entities representation $\bm{H}_{VE\gets T}$. We apply the average operation on the latter and then obtain the final representation of visual entities $\bm{x}_{ve}$. 

\textbf{MCT between Textual Features and Visual CNN Features.} Visual entities focus on the local high-level semantics of the images, while ignoring the global low-level visual features.
As a supplement, we use the multimodal co-attention transformer to model the correlations between textual features and visual CNN features.
Specifically,
we feed $\bm{H}_{T\gets VE}$ and the visual CNN features $\bm{H}_V$ into the second co-attention transformer, obtaining the textual representation enhanced by both visual entities and visual CNN features $\bm{H}_{T\gets (VE,V)}$ and text-enhanced visual representation $\bm{H}_{V\gets T}$.
We apply the average operation on the above features to obtain the final representation of the text and image, that is $\bm{x}_t$ and $\bm{x}_v$, respectively.

\subsubsection{Entity Inconsistency Measurement}
Multimodal entity inconsistency is a potential indicator for multimodal fake news. For example, if the person mentioned in the text is inconsistent with the recognized celebrity in the image, this news post may be fake with misused images (see Figure ~\ref{fig:outdated}).
Motivated by M{\"u}ller-Budack et al. \cite{icmrentity}, we measure the multimodal entity inconsistency of person, location, and a more general event context.
There are two challenges for this measurement: the first one is the heterogeneity of textual and visual features. 
Unlike previous works that calculate the multimodal similarity in transformed \cite{mcnn} or visual feature spaces\cite{icmrentity}, we calculate the similarity of multimodal entities in textual feature space based on their word embeddings.
Second, news text usually contains more entities and information than the companying images, and thus some textual entities could be without the aligned visual entities. 
Considering that fake news commonly tampers only one entity type to maintain credibility, we consider the multimodal news as entity inconsistent only when there are no aligned multimodal entities. 

Taking person entity as an example, we define the cross-modal person similarity as the maximum similarity among all pairs of textual and visual person entities. 
Since neural networks have inevitable errors when detecting visual entities, the confidence is considered when computing the similarity.  
Formally, 
we define $\bm{t}$ and $\bm{v}$ as the feature vectors of the textual and visual entities.
For a news post with $\bm{T}_p$ and $\bm{V}_p$, we calculate the cross-modal person similarity as 
\begin{equation}
	x_s^p=\max_{\bm{t} \in \bm{T}_p}(\sum_{\bm{v} \in \bm{V}_p} \rho(\bm{v}) \frac{\bm{t} \cdot \bm{v}}{\|\bm{t}\| \|\bm{v}\|}),
\end{equation}
where $\rho(\bm{v}) $ represents the probability of visual entity $\bm{v}$. 
For news that lacks textual or visual entities, we set the similarity as 1 to indicate no effective clue about multimodal inconsistency for fake news detection.  
Similarly, we compute the cross-modal location similarity $x_s^l$ and context similarity $x_s^c$, and then concatenate them to form the entity consistency feature $\bm{x}_s = [x_s^p, x_s^l, x_s^c]$. 

Finally, we concatenate the final representation of the text $\bm{x}_t$, that of visual entities $\bm{x}_{ve}$, that of the image $\bm{x}_{v}$, and the multimodal entity consistency feature $\bm{x}_s$ to obtain the final multimodal representation as
\begin{equation}
	\bm{x}_m = {\rm concat}(\bm{x}_t, \bm{x}_{ve}, \bm{x}_v, \bm{x}_s).
\end{equation}

\subsection{Classification}
\label{classification}
Till now, we have obtained the final multimodal representation $\bm{x}_m$, which models the input multimodal news from multiple perspectives.
We use a fully connected layer with softmax activation to project the multimodal feature vector $\bm{x}_m$ into the target space of two classes: real and fake news, and gain the probability distributions:
\begin{equation}
	\bm{p} = {\rm softmax}(\bm{W} \bm{x}_m + \bm{b}),
\end{equation}
where $\bm{p} = [p_0, p_1]$ is the predicted probability vector with $p_0$ and $p_1$ indicate the predicted probability of label being 0 (real news) and 1 (fake news), respectively. $\bm{W}$ is the weight matrix and $\bm{b}$ is the bias term. 
Thus, for each news post, the goal is to minimize the binary cross-entropy loss function as follows,
\begin{equation}
	\mathcal{L}_p=-[y \log p_0 + (1-y)\log p_1],
\end{equation}
where $y \in \{0,1\}$ denotes the ground-truth label.

\section{Experiments} 
In this section, we conduct experiments to evaluate the effectiveness of the proposed EM-FEND. 
Specifically, we aim to answer the following evaluation questions:
\begin{itemize}[leftmargin=17pt]
	\item \textbf{EQ1:} Can EM-FEND improve the classification performance of distinguishing multimodal fake and real news?
	\item \textbf{EQ2:} How effective are various visual features (especially the visual entities) and cross-modal correlations in improving the performance of EM-FEND?
	\item \textbf{EQ3:} How does EM-FEND perform in online fake news detection?
\end{itemize}

\subsection{Datasets}
To prove the generalization of the proposed EM-FEND, we conduct experiments on two real-world datasets of different languages. 
%from different social media platforms. 
%Table * shows the statistics of the two datasets. 
%The details are as follows:

\subsubsection{Chinese Dataset}
The Chinese dataset is constructed on the Chinese Sina Weibo microblogging platform by Jin et al. \cite{attRNN} and has been broadly used in existing works \cite{eann, mvae, spotfake}. 
The fake news posts are verified by the official rumor debunking website of Weibo\footnote{https://service.account.weibo.com}, which serves as a reputable source to collect fake news posts in literature. The real news posts are collected from Weibo during the same period as the fake news and are verified by Xinhua News Agency, an authoritative news agency in China. This dataset has been preprocessed to ensure that each post corresponds to an image. 
In total, this dataset includes 4,749 fake news posts and 4,779 real news posts with corresponding images. 

\subsubsection{English Dataset} The English dataset is proposed by Yang et al. \cite{ticnn}. 
The fake news is crawled from news websites that are manually assessed as low credibility\footnote{https://www.kaggle.com/mrisdal/fake-news}. And the real news is crawled from well-known authoritative news websites such as the New York Times.
After removing text-only news, non-English news, and news with unavailable images, 
we obtain 2,844 fake news articles and 2,825 real news articles, each corresponding to an image.

To prevent the model from overfitting on event topics, we first use the K-means algorithm to find the common events and split the data into training, validation and testing sets based on these event clusters to ensure that there is no event overlap among these sets \cite{eann}.
The training, validation, and testing sets contain data approximately with a ratio of 3:1:1. 
We use the Accuracy (Acc.) and Precision (Prec.), Recall and F1 score of the fake-news class as evaluation metrics.

\subsection{Implementation Details}
We use the pre-trained BERT models\footnote{https://github.com/google-research/bert} (i.e., bert-base-chinese and bert-base-uncased) to obtain the textual representation. The maximum sequence length is 256 for both datasets. 
For models that are not based on BERT, we use publicly available Word2Vec models\footnote{https://ai.tencent.com/ailab/nlp/en/embedding.html}\footnote{https://github.com/mmihaltz/word2vec-GoogleNews-vectors} to obtain the word embeddings.
For detecting textual entities, we use public API\footnote{https://ai.baidu.com/tech/nlp\_basic/lexical} and the open-sourced library Spacy\footnote{https://spacy.io/} for Chinese and English news, respectively.
In the co-attention transformer block, we employ 8 heads and the hidden size is set as 256 and 128 for EM-FEND and EM-FEND-base, respectively.  
The hidden size of LSTM in the EM-FEND-base is 128. 
We use a batch size of 64 instances in the training process.
The model is trained for 100 epochs with early stopping to prevent overfitting. We use ReLU as the non-linear activation function and use Adam\cite{adam} algorithm to optimize the loss function.
The dropout rate is set as 0.3.

\subsection{Comparison Methods}
\label{baseline}
To validate the effectiveness of the proposed EM-FEND framework, we compare it with several representative methods including single-modality and multimodal methods as follows:\\
\textbf{Single-modality Methods}
\begin{itemize}[leftmargin=17pt]
	\item \textbf{Bi-LSTM}: uses a network based on the bidirectional LSTM to classify the given piece of news. 
	\item \textbf{BERT}: uses a pre-trained BERT to obtain the representation of the given piece of news and a fully connected layer to make classifications. 
	\item \textbf{VGG19}: fine-tunes VGG19 to model news images for classifications. 
\end{itemize}
\textbf{Multimodal Methods}
\begin{itemize}[leftmargin=17pt]
	\item \textbf{attRNN-} \cite{attRNN}: proposes an innovative RNN with an attention mechanism for effectively fusing multimodal features. In detail, it produces the joint features of text and social context by an LSTM network and fuses them with visual features by utilizing the neural-level attention from the outputs of the LSTM. For a fair comparison, we remove the part dealing with social context features.
	\item \textbf{MVAE} \cite{mvae}: utilizes a multimodal variational autoencoder trained jointly with a fake news detector to learn a shared representation of multimodal content for fake news detection. It is composed of textual and visual encoders and corresponding decoders, and a fake news detector.
	\item \textbf{MKN} \cite{mkemn}: retrieves concepts of textual entities from external knowledge graphs and proposes a multi-channel word-knowledge-visual-aligned CNN for fusing multimodal information. The original model MKEMN uses an event memory network to detect fake news events. Because we focus on detecting fake news at the post level, we use the building block MKN of MKEMN, which deals with fake news posts, as a compared method. 
	\item \textbf{SAFE} \cite{safe}: translates the input image into a sentence, and computes the multimodal relevance based on the sentence similarity as the auxiliary loss for the fake news classification.  
	\item \textbf{SpotFake} \cite{spotfake}: concatenates the textual and visual features obtained from pre-trained BERT and VGG19 respectively for classification.
	\item \textbf{CARMN} \cite{ipm-song}: proposes a cross-modal attention residual network to fuse multimodal features. We use the pre-trained BERT to obtain the textual representation.   
\end{itemize}

Considering that using pre-trained language models to extract textual features usually improves the detection performance of models even without significant changes on the model structure \cite{bert}, we design a reduced variant of the proposed EM-FEND model to ensure the fairness of comparisons. 

\begin{itemize}[leftmargin=17pt]
	\item \textbf{EM-FEND-base}: uses a Bi-LSTM with pre-trained Word2Vec models to replace BERT in EM-FEND when obtaining the textual features. The embeddings of textual and visual entities are also obtained by pre-trained Word2Vec models.
\end{itemize}

\begin{table}
	\caption{Performance comparison for multimodal fake news detection on two real-world datasets.}
	\begin{center}
		\begin{tabular}{llrrrr}
			\hline
			& \textbf{Methods}&\textbf{Acc.}&\textbf{Prec.}&\textbf{Recall}&\textbf{F1}\\
			\hline
			\multirow{11}{*}{\textbf{\begin{sideways}Chinese\end{sideways}}} & Bi-LSTM & 0.785 & 0.851 & 0.692 & 0.763\\
			& BERT & 0.830 & \textbf{0.977} & 0.675 & 0.798 \\
			& VGG19 & 0.730 & 0.789 & 0.626 & 0.698 \\
			\cline{2-6}
			& attRNN-\cite{attRNN} & 0.808 & 0.882 & 0.711 & 0.787 \\
			& MVAE\cite{mvae} & 0.797 & 0.827 & 0.751 & 0.787\\
			& MKN\cite{mkemn} & 0.805 & 0.865 & 0.722 & 0.787\\
			& SAFE\cite{safe} & 0.790 & 0.886 & 0.665 & 0.760\\
			& EM-FEND-base (Ours) & 0.852 & 0.841 & \underline{0.853} & 0.847 \\
			\cline{2-6}
			& SpotFake\cite{spotfake} & 0.852 & 0.854 & 0.850 & \underline{0.852}\\
			& CARMN \cite{ipm-song} & \underline{0.865} & \underline{0.933} & 0.774 & 0.846\\
			& EM-FEND (Ours) & \textbf{0.904} & 0.897 & \textbf{0.904} & \textbf{0.901}\\
			\hline
			\multirow{11}{*}{\textbf{\begin{sideways}English\end{sideways}}}  & Bi-LSTM & 0.864 & 0.877 & 0.843 & 0.859\\
			& BERT & 0.873 & 0.869 & 0.875 & 0.872 \\
			& VGG19 & 0.773 & 0.783 & 0.747 & 0.764 \\
			\cline{2-6}
			& attRNN-\cite{attRNN} & 0.872 & 0.861 & 0.882 & 0.871 \\
			& MVAE\cite{mvae} & 0.879 & 0.902& 0.848 & 0.874\\
			& MKN\cite{mkemn} & 0.889 & 0.846 & 0.929 & 0.886\\
			& SAFE\cite{safe} & 0.909 & 0.922 & 0.890 & 0.906\\
			& EM-FEND-base (Ours) & \underline{0.943} & 0.926 & \underline{0.961} & \underline{0.943}\\
			\cline{2-6}
			& SpotFake\cite{spotfake} & 0.899 & 0.879 & 0.923 & 0.901 \\
			& CARMN \cite{ipm-song} & 0.937 & \underline{0.934}& 0.940& 0.937\\
			& EM-FEND (Ours) & \textbf{0.975} & \textbf{0.978} & \textbf{0.973} & \textbf{0.975} \\
			\hline			
		\end{tabular}
		\label{tab:main}
	\end{center}
\end{table}

\subsection{Performance Comparison (\textbf{EQ1})}

We compare EM-FEND with representative methods introduced in Section \ref{baseline}. The results are presented in Table ~\ref{tab:main}, from which we can draw the following observations:
\begin{itemize}[leftmargin=17pt]
	\item EM-FEND is much better than other methods on both datasets, no matter whether or not to adopt BERT as the textual feature extractor. It validates that EM-FEND can effectively capture important multimodal clues that existing works ignore to detect fake news. Specifically, EM-FEND and EM-FEND-base outperform the corresponding state-of-the-art methods by at least 3.8 and 3.4 percentage points in accuracy, respectively.
	\item Methods based on textual modality are better than the visual modality, proving that the text provides more rich clues than images. Multimodal methods are generally better than methods based on single-modality, indicating the complementarity of multimodal features.
	\item Pre-trained language models (i.e., BERT) can improve the performance of our method. It is mainly due to the strong ability of transformers in modeling context and the abundant knowledge injected in the pre-trained models.         
\end{itemize}

\subsection{Ablation Study (\textbf{EQ2})}
We design two groups of ablation experiments to evaluate the effectiveness of different components in EM-FEND.
Specifically, we design several internal models for comparison, which are simplified variations of EM-FEND with certain visual features removed:

\begin{itemize}[leftmargin=17pt]
	\item \textbf{w/o visual entities}: EM-FEND without the visual entities extraction, and the following co-attention transformer between textual features and visual entities and entity inconsistency measurement module.  
	\item \textbf{w/o OCR text}: EM-FEND without the OCR text.
	\item \textbf{w/o fine-tuned (FT) VGG feature}: We extract visual features from pre-trained VGG19 without fine-tuning.
\end{itemize}

Similarly, we design the following variants of EM-FEND to prove the effectiveness of different cross-modal correlations:
\begin{itemize}[leftmargin=17pt]
	\item \textbf{w/o co-attention-ve}: EM-FEND without the co-attention transformer between textual features and visual entities.
	\item \textbf{w/o co-attention-vf}: EM-FEND without the co-attention transformer between textual and visual CNN features.
	\item \textbf{w/o entity inconsistency measurement}: EM-FEND without the entity inconsistency measurement module. 
\end{itemize} 

The results of the ablation study are reported in Table ~\ref{tab:ablation1} and Table ~\ref{tab:ablation2}. We have the following observations:

\begin{table}
	\caption{Ablation study on various visual features.}
	\begin{center}
		\begin{tabular}{llrrrr}
			\hline
			& \textbf{Methods}&\textbf{Acc.}&\textbf{Prec.}&\textbf{Recall}&\textbf{F1}\\
			\hline
			\multirow{4}{*}{\textbf{\begin{sideways}Chinese\end{sideways}}} & EM-FEND & \textbf{0.904} & 0.897 & \textbf{0.904} & \textbf{0.901}\\
			& w/o visual entities & 0.886 & \textbf{0.930} & 0.823& 0.873\\
			& w/o OCR text & 0.882 & 0.902& 0.845& 0.873\\
			& w/o FT VGG feature & 0.773 & 0.783 & 0.747 & 0.764\\
			\hline
			\multirow{4}{*}{\textbf{\begin{sideways}English\end{sideways}}}  & EM-FEND & \textbf{0.975} & \textbf{0.978} & \textbf{0.973} & \textbf{0.975} \\
			& w/o visual entities & 0.953 & 0.954 & 0.950 & 0.952 \\
			& w/o OCR text & 0.970 & 0.967 & 0.972 & 0.969 \\
			& w/o FT VGG feature  & 0.970 & 0.954 & 0.988 & 0.971\\
			\hline			
		\end{tabular}
		\label{tab:ablation1}
	\end{center}
\end{table}

1) \textit{Visual Features:} 
All of these three visual features are important for fake news detection. However, the most important visual features on these two datasets are different: fine-tuned VGG features in the Chinese dataset and visual entities in the English dataset. 
This phenomenon results from the differences in data sources between these two datasets. 
The Chinese dataset is collected from the social media platform, and thus the multimodal fake news is more likely to show characteristics of low image quality brought by wide propagation. 
Differently, the English dataset originates from the formal news websites, of which the news has high-quality and informative images. Thus, the high-level visual semantic features are more important than low-level visual features for detecting this kind of multimodal fake news.   
This phenomenon also proves the generalization ability of EM-FEND in detecting different types of multimodal fake news.  

\begin{table}
	\caption{Ablation study on various cross-modal correlations.}
	\begin{center}
		\begin{tabular}{llrrrr}
			\hline
			&\textbf{Methods}&\textbf{Acc.}&\textbf{Prec.}&\textbf{Recall}&\textbf{F1}\\
			\hline
			\multirow{4}{*}{\textbf{\begin{sideways}Chinese\end{sideways}}} &  EM-FEND & \textbf{0.904} & 0.897 & \textbf{0.904} & \textbf{0.901}\\
			& w/o entity consistency & 0.899 & \textbf{0.932} & 0.849 & 0.889 \\
			& w/o co-attention-ve & 0.890 & 0.914 & 0.851 & 0.881 \\
			& w/o co-attention-vf & 0.886 & 0.901 & 0.855 & 0.878 \\
			\hline
			\multirow{4}{*}{\textbf{\begin{sideways}English\end{sideways}}} & EM-FEND & \textbf{0.975} & \textbf{0.978} & \textbf{0.973} & \textbf{0.975} \\
			& w/o entity consistency & 0.962 & 0.977 & 0.945 & 0.961 \\
			& w/o co-attention-ve & 0.959 & 0.953 & 0.966 & 0.959 \\
			& w/o co-attention-vf & 0.930 & 0.937 & 0.920 & 0.928 \\
			\hline			
		\end{tabular}
		\label{tab:ablation2}
	\end{center}
\end{table}

2) \textit{Cross-modal Correlations:}
Other than visual features, the various cross-modal correlations are also important for achieving the best performance of EM-FEND.
If we remove one of them, the performance will drop by a certain degree.
Specifically, 
i) the accuracy is lower than the complete model by at least 1.4 percentage points in accuracy when we replace the single co-attention transformer module with the average operation, proving that the co-attention transformer can effectively fuse multimodal features by capturing the multimodal alignment; 
ii) The influence of entity consistency is smaller than other cross-modal correlations, probably due to the sparsity of visual entities and the noises brought by entity detectors.

\subsection{Robustness to Imbalanced Online Data (\textbf{EQ3})}
In real-world scenarios, the number of fake news is much lower than real news, which means that the online news data that needs to be detected is unbalanced.
We collect news from an online fake news detection system like \cite{wwwsystem} during 9 months.
After removing news posts without text or images and duplicated posts, we obtain 217 multimodal fake news posts and 3353 real news posts annotated by experts, with a ratio of 1:15 approximately.   
It is worth noting that it's more difficult to distinguish these fake and real news than distinguishing that in datasets used in Section 4.1, because these real news posts originate from suspicious news and usually show typical patterns of fake news.       

\begin{figure}[htbp]
	\centering
\includegraphics[width=.35\textwidth]{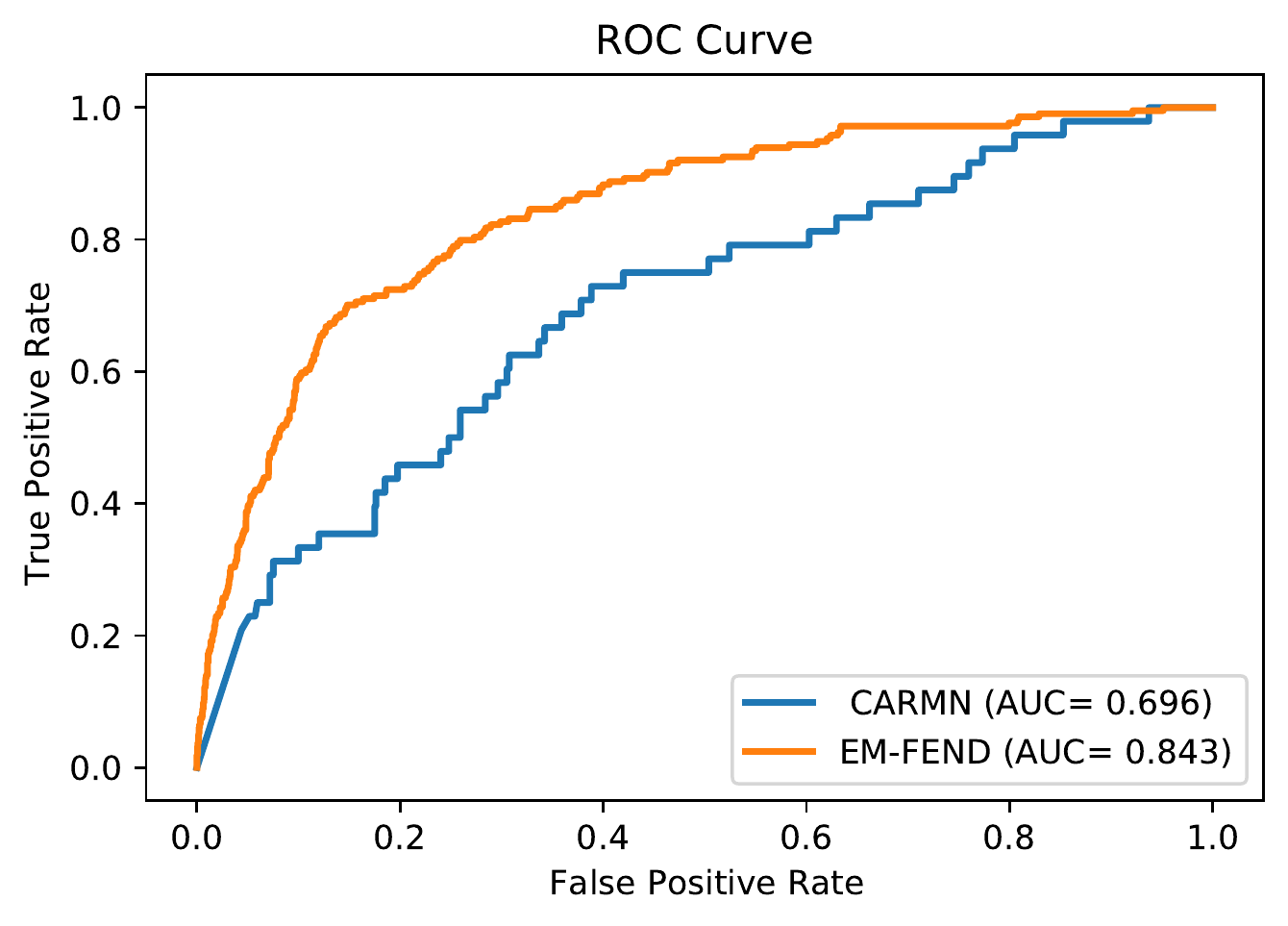}
	\caption{ROC curves of EM-FEND and CARMN.}
	\label{fig:online}
\end{figure}

To evaluate the robustness of EM-FEND to imbalanced online data, we compare EM-FEND with CARMN \cite{ipm-song}, the best competitor to EM-FEND (Table ~\ref{tab:main}), in the imbalanced dataset. 
Figure ~\ref{fig:online} shows the ROC curves of these two models, from which we observe that EM-FEND outperforms CARMN in online data.

\subsection{Case Study}
\begin{figure}[htbp]
	\centering
	\includegraphics[height=1.8in]{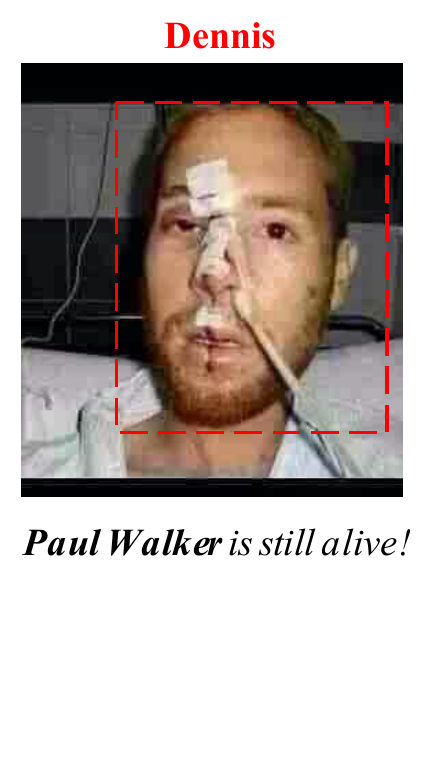}
	\includegraphics[height=1.8in]{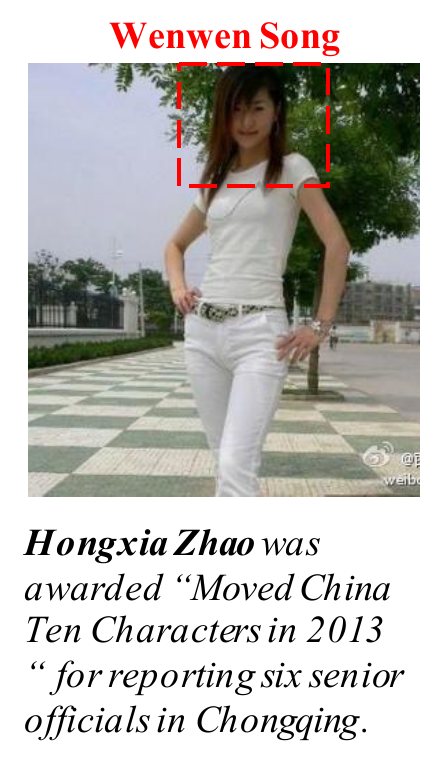}
	\includegraphics[height=1.8in]{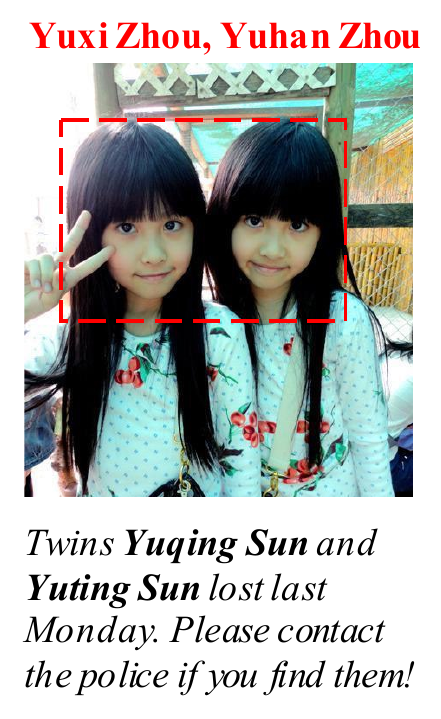}
	\caption{Some fake news with low multimodal person consistency. In these cases, the person entity mentioned in the text is inconsistent with that recognized in the image.}
	\label{fig:entity}
\end{figure}

In this part, we show some cases to intuitively show the behaviors of entity inconsistency measurement module in EM-FEND. Specifically, we list several representative multimodal fake news that are measured as low person consistency in Figure ~\ref{fig:entity}.
It shows that this module can effectively measure the multimodal entity inconsistency as easily understanding explanations for the model's decisions about fake news.

\section{Conclusion}
In this paper, we find three valuable cross-modal correlations in multimodal fake news on social media, that is entity inconsistency, mutual enhancement and text complementation, which provides diverse multimodal clues. Also, we reveal the importance of visual entities in understanding news-related visual semantics and capturing these multimodal clues.  
Accordingly, we propose a novel entity-enhanced multimodal fusion framework named EM-FEND to simultaneously model three cross-modal correlations. 
Extensive experiments have proved the effectiveness of EM-FEND.

%%
%% The acknowledgments section is defined using the "acks" environment
%% (and NOT an unnumbered section). This ensures the proper
%% identification of the section in the article metadata, and the
%% consistent spelling of the heading.
\begin{acks}
This work was supported by the National Key Research and Development Program of China (2017YFC0820604), and the National Natural Science Foundation of China (U1703261, 62172420).
%The corresponding author is Juan Cao.
\end{acks}

%%
%% The next two lines define the bibliography style to be used, and
%% the bibliography file.
%\vfill\eject
\bibliographystyle{ACM-Reference-Format}
\balance
\bibliography{sample-base}

%%% -*-BibTeX-*-
%%% Do NOT edit. File created by BibTeX with style
%%% ACM-Reference-Format-Journals [18-Jan-2012].

\begin{thebibliography}{35}

%%% ====================================================================
%%% NOTE TO THE USER: you can override these defaults by providing
%%% customized versions of any of these macros before the \bibliography
%%% command.  Each of them MUST provide its own final punctuation,
%%% except for \shownote{}, \showDOI{}, and \showURL{}.  The latter two
%%% do not use final punctuation, in order to avoid confusing it with
%%% the Web address.
%%%
%%% To suppress output of a particular field, define its macro to expand
%%% to an empty string, or better, \unskip, like this:
%%%
%%% \newcommand{\showDOI}[1]{\unskip}   % LaTeX syntax
%%%
%%% \def \showDOI #1{\unskip}           % plain TeX syntax
%%%
%%% ====================================================================

\ifx \showCODEN    \undefined \def \showCODEN     #1{\unskip}     \fi
\ifx \showDOI      \undefined \def \showDOI       #1{#1}\fi
\ifx \showISBNx    \undefined \def \showISBNx     #1{\unskip}     \fi
\ifx \showISBNxiii \undefined \def \showISBNxiii  #1{\unskip}     \fi
\ifx \showISSN     \undefined \def \showISSN      #1{\unskip}     \fi
\ifx \showLCCN     \undefined \def \showLCCN      #1{\unskip}     \fi
\ifx \shownote     \undefined \def \shownote      #1{#1}          \fi
\ifx \showarticletitle \undefined \def \showarticletitle #1{#1}   \fi
\ifx \showURL      \undefined \def \showURL       {\relax}        \fi
% The following commands are used for tagged output and should be
% invisible to TeX
\providecommand\bibfield[2]{#2}
\providecommand\bibinfo[2]{#2}
\providecommand\natexlab[1]{#1}
\providecommand\showeprint[2][]{arXiv:#2}

\bibitem[\protect\citeauthoryear{Boididou, Andreadou, Papadopoulos,
  Dang-Nguyen, Boato, Riegler, Kompatsiaris, et~al\mbox{.}}{Boididou
  et~al\mbox{.}}{2015}]%
        {mediaeval15}
\bibfield{author}{\bibinfo{person}{Christina Boididou},
  \bibinfo{person}{Katerina Andreadou}, \bibinfo{person}{Symeon Papadopoulos},
  \bibinfo{person}{Duc-Tien Dang-Nguyen}, \bibinfo{person}{Giulia Boato},
  \bibinfo{person}{Michael Riegler}, \bibinfo{person}{Yiannis Kompatsiaris},
  {et~al\mbox{.}}} \bibinfo{year}{2015}\natexlab{}.
\newblock \showarticletitle{Verifying Multimedia Use at MediaEval 2015}. In
  \bibinfo{booktitle}{\emph{Working Notes Proceedings of the MediaEval 2015
  Workshop}}.
\newblock


\bibitem[\protect\citeauthoryear{Boididou, Papadopoulos, Dang-Nguyen, Boato,
  Riegler, Middleton, Petlund, Kompatsiaris, et~al\mbox{.}}{Boididou
  et~al\mbox{.}}{2016}]%
        {mediaeval16}
\bibfield{author}{\bibinfo{person}{Christina Boididou}, \bibinfo{person}{Symeon
  Papadopoulos}, \bibinfo{person}{Duc-Tien Dang-Nguyen},
  \bibinfo{person}{Giulia Boato}, \bibinfo{person}{Michael Riegler},
  \bibinfo{person}{Stuart~E. Middleton}, \bibinfo{person}{Andreas Petlund},
  \bibinfo{person}{Yiannis Kompatsiaris}, {et~al\mbox{.}}}
  \bibinfo{year}{2016}\natexlab{}.
\newblock \showarticletitle{Verifying Multimedia Use at MediaEval 2016}. In
  \bibinfo{booktitle}{\emph{Working Notes Proceedings of the MediaEval 2016
  Workshop}}.
\newblock


\bibitem[\protect\citeauthoryear{Cao, Qi, Sheng, Yang, Guo, and Li}{Cao
  et~al\mbox{.}}{2020}]%
        {chapter}
\bibfield{author}{\bibinfo{person}{Juan Cao}, \bibinfo{person}{Peng Qi},
  \bibinfo{person}{Qiang Sheng}, \bibinfo{person}{Tianyun Yang},
  \bibinfo{person}{Junbo Guo}, {and} \bibinfo{person}{Jintao Li}.}
  \bibinfo{year}{2020}\natexlab{}.
\newblock \showarticletitle{Exploring the Role of Visual Content in Fake News
  Detection}.
\newblock \bibinfo{journal}{\emph{Disinformation, Misinformation, and Fake News
  in Social Media}} (\bibinfo{year}{2020}), \bibinfo{pages}{141--161}.
\newblock


\bibitem[\protect\citeauthoryear{Castillo, Mendoza, and Poblete}{Castillo
  et~al\mbox{.}}{2011}]%
        {textCastillo}
\bibfield{author}{\bibinfo{person}{Carlos Castillo}, \bibinfo{person}{Marcelo
  Mendoza}, {and} \bibinfo{person}{Barbara Poblete}.}
  \bibinfo{year}{2011}\natexlab{}.
\newblock \showarticletitle{Information Credibility on Twitter}. In
  \bibinfo{booktitle}{\emph{Proceedings of the 20th International Conference on
  World Wide Web}}. \bibinfo{pages}{675--684}.
\newblock


\bibitem[\protect\citeauthoryear{Deng, Dong, Socher, Li, Li, and Fei-Fei}{Deng
  et~al\mbox{.}}{2009}]%
        {imagenet}
\bibfield{author}{\bibinfo{person}{Jia Deng}, \bibinfo{person}{Wei Dong},
  \bibinfo{person}{Richard Socher}, \bibinfo{person}{Li-Jia Li},
  \bibinfo{person}{Kai Li}, {and} \bibinfo{person}{Li Fei-Fei}.}
  \bibinfo{year}{2009}\natexlab{}.
\newblock \showarticletitle{ImageNet: A large-scale Hierarchical Image
  Database}. In \bibinfo{booktitle}{\emph{2009 {IEEE} Computer Society
  Conference on Computer Vision and Pattern Recognition}}.
  \bibinfo{pages}{248--255}.
\newblock


\bibitem[\protect\citeauthoryear{Devlin, Chang, Lee, and Toutanova}{Devlin
  et~al\mbox{.}}{2019}]%
        {bert}
\bibfield{author}{\bibinfo{person}{Jacob Devlin}, \bibinfo{person}{Ming-Wei
  Chang}, \bibinfo{person}{Kenton Lee}, {and} \bibinfo{person}{Kristina
  Toutanova}.} \bibinfo{year}{2019}\natexlab{}.
\newblock \showarticletitle{BERT: Pre-training of Deep Bidirectional
  Transformers for Language Understanding}. In
  \bibinfo{booktitle}{\emph{Proceedings of the 2019 Conference of the North
  American Chapter of the Association for Computational Linguistics: Human
  Language Technologies}}. \bibinfo{pages}{4171--4186}.
\newblock


\bibitem[\protect\citeauthoryear{Dhruv, Jaipal~Singh, Manish, and
  Vasudeva}{Dhruv et~al\mbox{.}}{2019}]%
        {mvae}
\bibfield{author}{\bibinfo{person}{Khattar Dhruv}, \bibinfo{person}{Goud
  Jaipal~Singh}, \bibinfo{person}{Gupta Manish}, {and} \bibinfo{person}{Varma
  Vasudeva}.} \bibinfo{year}{2019}\natexlab{}.
\newblock \showarticletitle{MVAE: Multimodal Variational Autoencoder for Fake
  News Detection}. In \bibinfo{booktitle}{\emph{The World Wide Web
  Conference}}. \bibinfo{pages}{2915--2921}.
\newblock


\bibitem[\protect\citeauthoryear{Guo, Ding, Yao, Liang, and Yu}{Guo
  et~al\mbox{.}}{2020}]%
        {surveycsur2020}
\bibfield{author}{\bibinfo{person}{Bin Guo}, \bibinfo{person}{Yasan Ding},
  \bibinfo{person}{Lina Yao}, \bibinfo{person}{Yunji Liang}, {and}
  \bibinfo{person}{Zhiwen Yu}.} \bibinfo{year}{2020}\natexlab{}.
\newblock \showarticletitle{The Future of False Information Detection on Social
  Media: New Perspectives and Trends}.
\newblock \bibinfo{journal}{\emph{Comput. Surveys}} \bibinfo{volume}{53},
  \bibinfo{number}{4} (\bibinfo{year}{2020}), \bibinfo{pages}{1--36}.
\newblock


\bibitem[\protect\citeauthoryear{Jin, Cao, Guo, Zhang, and Luo}{Jin
  et~al\mbox{.}}{2017}]%
        {attRNN}
\bibfield{author}{\bibinfo{person}{Zhiwei Jin}, \bibinfo{person}{Juan Cao},
  \bibinfo{person}{Han Guo}, \bibinfo{person}{Yongdong Zhang}, {and}
  \bibinfo{person}{Jiebo Luo}.} \bibinfo{year}{2017}\natexlab{}.
\newblock \showarticletitle{Multimodal Fusion with Recurrent Neural Networks
  for Rumor Detection on Microblogs}. In \bibinfo{booktitle}{\emph{Proceedings
  of the 25th ACM International Conference on Multimedia}}.
  \bibinfo{pages}{795--816}.
\newblock


\bibitem[\protect\citeauthoryear{Kingma and Ba}{Kingma and Ba}{2015}]%
        {adam}
\bibfield{author}{\bibinfo{person}{Diederik~P Kingma} {and}
  \bibinfo{person}{Jimmy Ba}.} \bibinfo{year}{2015}\natexlab{}.
\newblock \showarticletitle{Adam: A Method for Stochastic Optimization}. In
  \bibinfo{booktitle}{\emph{3rd International Conference on Learning
  Representations}}.
\newblock


\bibitem[\protect\citeauthoryear{Kumar and Shah}{Kumar and Shah}{2018}]%
        {surveykumar2018false}
\bibfield{author}{\bibinfo{person}{Srijan Kumar} {and} \bibinfo{person}{Neil
  Shah}.} \bibinfo{year}{2018}\natexlab{}.
\newblock \showarticletitle{False Information on Web and Social Media: A
  Survey}.
\newblock \bibinfo{journal}{\emph{arXiv preprint arXiv:1804.08559}}
  (\bibinfo{year}{2018}).
\newblock


\bibitem[\protect\citeauthoryear{Li, Sun, Yu, Tian, Yao, and Xu}{Li
  et~al\mbox{.}}{2021}]%
        {emaf}
\bibfield{author}{\bibinfo{person}{Peiguang Li}, \bibinfo{person}{Xian Sun},
  \bibinfo{person}{Hongfeng Yu}, \bibinfo{person}{Yu Tian},
  \bibinfo{person}{Fanglong Yao}, {and} \bibinfo{person}{Guangluan Xu}.}
  \bibinfo{year}{2021}\natexlab{}.
\newblock \showarticletitle{Entity-Oriented Multi-Modal Alignment and Fusion
  Network for Fake News Detection}.
\newblock \bibinfo{journal}{\emph{IEEE Transactions on Multimedia}}
  (\bibinfo{year}{2021}).
\newblock


\bibitem[\protect\citeauthoryear{Lu, Batra, Parikh, and Lee}{Lu
  et~al\mbox{.}}{2019}]%
        {vilbert}
\bibfield{author}{\bibinfo{person}{Jiasen Lu}, \bibinfo{person}{Dhruv Batra},
  \bibinfo{person}{Devi Parikh}, {and} \bibinfo{person}{Stefan Lee}.}
  \bibinfo{year}{2019}\natexlab{}.
\newblock \showarticletitle{ViLBERT: Pretraining Task-Agnostic Visiolinguistic
  Representations for Vision-and-Language Tasks}. In
  \bibinfo{booktitle}{\emph{Advances in Neural Information Processing
  Systems}}. \bibinfo{pages}{13--23}.
\newblock


\bibitem[\protect\citeauthoryear{Lu, Yang, Batra, and Parikh}{Lu
  et~al\mbox{.}}{2016}]%
        {2coattention}
\bibfield{author}{\bibinfo{person}{Jiasen Lu}, \bibinfo{person}{Jianwei Yang},
  \bibinfo{person}{Dhruv Batra}, {and} \bibinfo{person}{Devi Parikh}.}
  \bibinfo{year}{2016}\natexlab{}.
\newblock \showarticletitle{Hierarchical Question-Image Co-Attention for Visual
  Question Answering}. In \bibinfo{booktitle}{\emph{Proceedings of the 30th
  International Conference on Neural Information Processing Systems}}.
  \bibinfo{pages}{289--297}.
\newblock


\bibitem[\protect\citeauthoryear{Ma, Gao, Mitra, Kwon, Jansen, Wong, and
  Cha}{Ma et~al\mbox{.}}{2016}]%
        {majingijcai}
\bibfield{author}{\bibinfo{person}{Jing Ma}, \bibinfo{person}{Wei Gao},
  \bibinfo{person}{Prasenjit Mitra}, \bibinfo{person}{Sejeong Kwon},
  \bibinfo{person}{Bernard~J Jansen}, \bibinfo{person}{Kam-Fai Wong}, {and}
  \bibinfo{person}{Meeyoung Cha}.} \bibinfo{year}{2016}\natexlab{}.
\newblock \showarticletitle{Detecting Rumors from Microblogs with Recurrent
  Neural Networks}. In \bibinfo{booktitle}{\emph{Proceedings of the
  Twenty-Fifth International Joint Conference on Artificial Intelligence}}.
  \bibinfo{pages}{3818--3824}.
\newblock


\bibitem[\protect\citeauthoryear{M{\"u}ller-Budack, Theiner, Diering, Idahl,
  and Ewerth}{M{\"u}ller-Budack et~al\mbox{.}}{2020}]%
        {icmrentity}
\bibfield{author}{\bibinfo{person}{Eric M{\"u}ller-Budack},
  \bibinfo{person}{Jonas Theiner}, \bibinfo{person}{Sebastian Diering},
  \bibinfo{person}{Maximilian Idahl}, {and} \bibinfo{person}{Ralph Ewerth}.}
  \bibinfo{year}{2020}\natexlab{}.
\newblock \showarticletitle{Multimodal Analytics for Real-world News Using
  Measures of Cross-modal Entity Consistency}. In
  \bibinfo{booktitle}{\emph{Proceedings of the 2020 International Conference on
  Multimedia Retrieval}}. \bibinfo{pages}{16--25}.
\newblock


\bibitem[\protect\citeauthoryear{Nadeau and Sekine}{Nadeau and Sekine}{2007}]%
        {namedentity}
\bibfield{author}{\bibinfo{person}{David Nadeau} {and} \bibinfo{person}{Satoshi
  Sekine}.} \bibinfo{year}{2007}\natexlab{}.
\newblock \showarticletitle{A Survey of Named Entity Recognition and
  Classification}.
\newblock \bibinfo{journal}{\emph{Lingvisticae Investigationes}}
  \bibinfo{volume}{30}, \bibinfo{number}{1} (\bibinfo{year}{2007}),
  \bibinfo{pages}{3--26}.
\newblock


\bibitem[\protect\citeauthoryear{P{\'e}rez-Rosas, Kleinberg, Lefevre, and
  Mihalcea}{P{\'e}rez-Rosas et~al\mbox{.}}{2018}]%
        {texticcl2018}
\bibfield{author}{\bibinfo{person}{Ver{\'o}nica P{\'e}rez-Rosas},
  \bibinfo{person}{Bennett Kleinberg}, \bibinfo{person}{Alexandra Lefevre},
  {and} \bibinfo{person}{Rada Mihalcea}.} \bibinfo{year}{2018}\natexlab{}.
\newblock \showarticletitle{Automatic Detection of Fake News}. In
  \bibinfo{booktitle}{\emph{Proceedings of the 27th International Conference on
  Computational Linguistics}}. \bibinfo{pages}{3391--3401}.
\newblock


\bibitem[\protect\citeauthoryear{Qazvinian, Rosengren, Radev, and
  Mei}{Qazvinian et~al\mbox{.}}{2011}]%
        {textemnlp2011}
\bibfield{author}{\bibinfo{person}{Vahed Qazvinian}, \bibinfo{person}{Emily
  Rosengren}, \bibinfo{person}{Dragomir Radev}, {and} \bibinfo{person}{Qiaozhu
  Mei}.} \bibinfo{year}{2011}\natexlab{}.
\newblock \showarticletitle{Rumor has it: Identifying Misinformation in
  Microblogs}. In \bibinfo{booktitle}{\emph{Proceedings of the 2011 Conference
  on Empirical Methods in Natural Language Processing}}.
  \bibinfo{pages}{1589--1599}.
\newblock


\bibitem[\protect\citeauthoryear{Qi, Cao, Yang, Guo, and Li}{Qi
  et~al\mbox{.}}{2019}]%
        {mvnn}
\bibfield{author}{\bibinfo{person}{Peng Qi}, \bibinfo{person}{Juan Cao},
  \bibinfo{person}{Tianyun Yang}, \bibinfo{person}{Junbo Guo}, {and}
  \bibinfo{person}{Jintao Li}.} \bibinfo{year}{2019}\natexlab{}.
\newblock \showarticletitle{Exploiting Multi-domain Visual Information for Fake
  News Detection}. In \bibinfo{booktitle}{\emph{{IEEE} International Conference
  on Data Mining}}. \bibinfo{pages}{518--527}.
\newblock


\bibitem[\protect\citeauthoryear{Shu, Sliva, Wang, Tang, and Liu}{Shu
  et~al\mbox{.}}{2017}]%
        {surveykdd2017}
\bibfield{author}{\bibinfo{person}{Kai Shu}, \bibinfo{person}{Amy Sliva},
  \bibinfo{person}{Suhang Wang}, \bibinfo{person}{Jiliang Tang}, {and}
  \bibinfo{person}{Huan Liu}.} \bibinfo{year}{2017}\natexlab{}.
\newblock \showarticletitle{Fake News Detection on Social Media: A Data Mining
  Perspective}.
\newblock \bibinfo{journal}{\emph{ACM SIGKDD Explorations Newsletter}}
  \bibinfo{volume}{19}, \bibinfo{number}{1} (\bibinfo{year}{2017}),
  \bibinfo{pages}{22--36}.
\newblock


\bibitem[\protect\citeauthoryear{Simonyan and Zisserman}{Simonyan and
  Zisserman}{2015}]%
        {vgg19}
\bibfield{author}{\bibinfo{person}{Karen Simonyan} {and}
  \bibinfo{person}{Andrew Zisserman}.} \bibinfo{year}{2015}\natexlab{}.
\newblock \showarticletitle{Very Deep Convolutional Networks for Large-Scale
  Image Recognition}. In \bibinfo{booktitle}{\emph{3rd International Conference
  on Learning Representations}}.
\newblock


\bibitem[\protect\citeauthoryear{Singhal, Shah, Chakraborty, Kumaraguru, and
  Satoh}{Singhal et~al\mbox{.}}{2019}]%
        {spotfake}
\bibfield{author}{\bibinfo{person}{Shivangi Singhal},
  \bibinfo{person}{Rajiv~Ratn Shah}, \bibinfo{person}{Tanmoy Chakraborty},
  \bibinfo{person}{Ponnurangam Kumaraguru}, {and} \bibinfo{person}{Shin'ichi
  Satoh}.} \bibinfo{year}{2019}\natexlab{}.
\newblock \showarticletitle{SpotFake: A Multi-modal Framework for Fake News
  Detection}. In \bibinfo{booktitle}{\emph{Fifth {IEEE} International
  Conference on Multimedia Big Data}}. \bibinfo{pages}{39--47}.
\newblock


\bibitem[\protect\citeauthoryear{Song, Ning, Zhang, and Wu}{Song
  et~al\mbox{.}}{2021}]%
        {ipm-song}
\bibfield{author}{\bibinfo{person}{Chenguang Song}, \bibinfo{person}{Nianwen
  Ning}, \bibinfo{person}{Yunlei Zhang}, {and} \bibinfo{person}{Bin Wu}.}
  \bibinfo{year}{2021}\natexlab{}.
\newblock \showarticletitle{A Multimodal Fake News Detection Model Based on
  Crossmodal Attention Residual and Multichannel Convolutional Neural
  Networks}.
\newblock \bibinfo{journal}{\emph{Information Processing \& Management}}
  \bibinfo{volume}{58}, \bibinfo{number}{1} (\bibinfo{year}{2021}),
  \bibinfo{pages}{102437}.
\newblock


\bibitem[\protect\citeauthoryear{Vaswani, Shazeer, Parmar, Uszkoreit, Jones,
  Gomez, Kaiser, and Polosukhin}{Vaswani et~al\mbox{.}}{2017}]%
        {transformer}
\bibfield{author}{\bibinfo{person}{Ashish Vaswani}, \bibinfo{person}{Noam
  Shazeer}, \bibinfo{person}{Niki Parmar}, \bibinfo{person}{Jakob Uszkoreit},
  \bibinfo{person}{Llion Jones}, \bibinfo{person}{Aidan~N Gomez},
  \bibinfo{person}{{\L}ukasz Kaiser}, {and} \bibinfo{person}{Illia
  Polosukhin}.} \bibinfo{year}{2017}\natexlab{}.
\newblock \showarticletitle{Attention Is All You Need}. In
  \bibinfo{booktitle}{\emph{Advances in Neural Information Processing
  Systems}}. \bibinfo{pages}{5998--6008}.
\newblock


\bibitem[\protect\citeauthoryear{Wang, Ma, Jin, Yuan, Xun, Jha, Su, and
  Gao}{Wang et~al\mbox{.}}{2018}]%
        {eann}
\bibfield{author}{\bibinfo{person}{Yaqing Wang}, \bibinfo{person}{Fenglong Ma},
  \bibinfo{person}{Zhiwei Jin}, \bibinfo{person}{Ye Yuan},
  \bibinfo{person}{Guangxu Xun}, \bibinfo{person}{Kishlay Jha},
  \bibinfo{person}{Lu Su}, {and} \bibinfo{person}{Jing Gao}.}
  \bibinfo{year}{2018}\natexlab{}.
\newblock \showarticletitle{EANN: Event Adversarial Neural Networks for
  Multi-Modal Fake News Detection}. In \bibinfo{booktitle}{\emph{Proceedings of
  the 24th {ACM} {SIGKDD} International Conference on Knowledge Discovery {\&}
  Data Mining}}. \bibinfo{pages}{849--857}.
\newblock


\bibitem[\protect\citeauthoryear{Wang, Ma, Wang, Jha, and Gao}{Wang
  et~al\mbox{.}}{2021}]%
        {metaFEND}
\bibfield{author}{\bibinfo{person}{Yaqing Wang}, \bibinfo{person}{Fenglong Ma},
  \bibinfo{person}{Haoyu Wang}, \bibinfo{person}{Kishlay Jha}, {and}
  \bibinfo{person}{Jing Gao}.} \bibinfo{year}{2021}\natexlab{}.
\newblock \showarticletitle{Multi-modal Emergent Fake News Detection via Meta
  Neural Process Networks}. In \bibinfo{booktitle}{\emph{Proceedings of the
  27th {ACM} {SIGKDD} International Conference on Knowledge Discovery {\&} Data
  Mining}}.
\newblock


\bibitem[\protect\citeauthoryear{Wang, Qian, Hu, Fang, and Xu}{Wang
  et~al\mbox{.}}{2020}]%
        {kmgcn}
\bibfield{author}{\bibinfo{person}{Youze Wang}, \bibinfo{person}{Shengsheng
  Qian}, \bibinfo{person}{Jun Hu}, \bibinfo{person}{Quan Fang}, {and}
  \bibinfo{person}{Changsheng Xu}.} \bibinfo{year}{2020}\natexlab{}.
\newblock \showarticletitle{Fake News Detection via Knowledge-Driven Multimodal
  Graph Convolutional Networks}. In \bibinfo{booktitle}{\emph{Proceedings of
  the 2020 International Conference on Multimedia Retrieval}}.
  \bibinfo{pages}{540--547}.
\newblock


\bibitem[\protect\citeauthoryear{Xue, Wang, Tian, Li, Shi, and Wei}{Xue
  et~al\mbox{.}}{2021}]%
        {mcnn}
\bibfield{author}{\bibinfo{person}{Junxiao Xue}, \bibinfo{person}{Yabo Wang},
  \bibinfo{person}{Yichen Tian}, \bibinfo{person}{Yafei Li},
  \bibinfo{person}{Lei Shi}, {and} \bibinfo{person}{Lin Wei}.}
  \bibinfo{year}{2021}\natexlab{}.
\newblock \showarticletitle{Detecting Fake News by Exploring the Consistency of
  Multimodal Data}.
\newblock \bibinfo{journal}{\emph{Information Processing and Management}}
  \bibinfo{volume}{58}, \bibinfo{number}{5} (\bibinfo{year}{2021}),
  \bibinfo{pages}{102610}.
\newblock


\bibitem[\protect\citeauthoryear{Yang, Zheng, Zhang, Cui, Li, and Yu}{Yang
  et~al\mbox{.}}{2018}]%
        {ticnn}
\bibfield{author}{\bibinfo{person}{Yang Yang}, \bibinfo{person}{Lei Zheng},
  \bibinfo{person}{Jiawei Zhang}, \bibinfo{person}{Qingcai Cui},
  \bibinfo{person}{Zhoujun Li}, {and} \bibinfo{person}{Philip~S Yu}.}
  \bibinfo{year}{2018}\natexlab{}.
\newblock \showarticletitle{TI-CNN: Convolutional Neural Networks for Fake News
  Detection}.
\newblock \bibinfo{journal}{\emph{arXiv preprint arXiv:1806.00749}}
  (\bibinfo{year}{2018}).
\newblock


\bibitem[\protect\citeauthoryear{Zafarani, Zhou, Shu, and Liu}{Zafarani
  et~al\mbox{.}}{2019}]%
        {surveykdd2019}
\bibfield{author}{\bibinfo{person}{Reza Zafarani}, \bibinfo{person}{Xinyi
  Zhou}, \bibinfo{person}{Kai Shu}, {and} \bibinfo{person}{Huan Liu}.}
  \bibinfo{year}{2019}\natexlab{}.
\newblock \showarticletitle{Fake News Research: Theories, Detection Strategies,
  and Open Problems}. In \bibinfo{booktitle}{\emph{Proceedings of the 25th
  {ACM} {SIGKDD} International Conference on Knowledge Discovery {\&} Data
  Mining}}. \bibinfo{pages}{3207--3208}.
\newblock


\bibitem[\protect\citeauthoryear{Zhang, Fang, Qian, and Xu}{Zhang
  et~al\mbox{.}}{2019}]%
        {mkemn}
\bibfield{author}{\bibinfo{person}{Huaiwen Zhang}, \bibinfo{person}{Quan Fang},
  \bibinfo{person}{Shengsheng Qian}, {and} \bibinfo{person}{Changsheng Xu}.}
  \bibinfo{year}{2019}\natexlab{}.
\newblock \showarticletitle{Multi-modal Knowledge-aware Event Memory Network
  for Social Media Rumor Detection}. In \bibinfo{booktitle}{\emph{Proceedings
  of the 27th ACM International Conference on Multimedia}}.
  \bibinfo{pages}{1942--1951}.
\newblock


\bibitem[\protect\citeauthoryear{Zhou, Cao, Jin, Xie, Su, Chu, Cao, and
  Zhang}{Zhou et~al\mbox{.}}{2015}]%
        {wwwsystem}
\bibfield{author}{\bibinfo{person}{Xing Zhou}, \bibinfo{person}{Juan Cao},
  \bibinfo{person}{Zhiwei Jin}, \bibinfo{person}{Fei Xie}, \bibinfo{person}{Yu
  Su}, \bibinfo{person}{Dafeng Chu}, \bibinfo{person}{Xuehui Cao}, {and}
  \bibinfo{person}{Junqiang Zhang}.} \bibinfo{year}{2015}\natexlab{}.
\newblock \showarticletitle{Real-time News Certification System on Sina Weibo}.
  In \bibinfo{booktitle}{\emph{Proceedings of the 24th International Conference
  on World Wide Web}}. \bibinfo{pages}{983--988}.
\newblock


\bibitem[\protect\citeauthoryear{Zhou, Wu, and Zafarani}{Zhou
  et~al\mbox{.}}{2020}]%
        {safe}
\bibfield{author}{\bibinfo{person}{Xinyi Zhou}, \bibinfo{person}{Jindi Wu},
  {and} \bibinfo{person}{Reza Zafarani}.} \bibinfo{year}{2020}\natexlab{}.
\newblock \showarticletitle{SAFE: Similarity-Aware Multi-modal Fake News
  Detection}. In \bibinfo{booktitle}{\emph{Pacific-Asia Conference on Knowledge
  Discovery and Data Mining}}. \bibinfo{pages}{354--367}.
\newblock


\bibitem[\protect\citeauthoryear{Zubiaga, Aker, Bontcheva, Liakata, and
  Procter}{Zubiaga et~al\mbox{.}}{2018}]%
        {surveycsur2018}
\bibfield{author}{\bibinfo{person}{Arkaitz Zubiaga}, \bibinfo{person}{Ahmet
  Aker}, \bibinfo{person}{Kalina Bontcheva}, \bibinfo{person}{Maria Liakata},
  {and} \bibinfo{person}{Rob Procter}.} \bibinfo{year}{2018}\natexlab{}.
\newblock \showarticletitle{Detection and Resolution of Rumours in Social
  Media: A Survey}.
\newblock \bibinfo{journal}{\emph{Comput. Surveys}} \bibinfo{volume}{51},
  \bibinfo{number}{2} (\bibinfo{year}{2018}), \bibinfo{pages}{32}.
\newblock


\end{thebibliography}

%%
%% If your work has an appendix, this is the place to put it.
%\appendix
%
%\section{Research Methods}

\end{document}